# ZYGOTIC COMBINATORIAL PROCESS IN PLANTS

E.V. Levites, S.S. Kirikovich

*Federal State-Maintained Institution of Science Institute of Cytology and Genetics, Siberian Branch of the Russian Academy of Sciences, Novosibirsk 630090; fax-333-12-78; e-mail:* levites@bionet.nsc.ru

Experimental data that prove the existence of the zygotic combinatorial process occurring in an embryogenesis-entering zygote are presented in the paper. The zygotic combinatorial process is found when analyzing $F_1$ hybrid plants obtained from crossing homozygous forms different, minimum, in two marker enzymes, and it is found in that hybrid plant which, with one marker enzyme heterozygous spectrum, has a homozygous spectrum of the other. The zygotic combinatorial process leads to $F_1$ hybrids uniformity aberration. The zygotic combinatory process revealed in the study is supposed to be conditioned by chromosome polyteny in mother plant cells and diminution of chromatin excess from the embryogenesis-entering zygote. An obligatory condition for combinatorial process is the presence of free exchange of cromatides among homological chromosomes in an embryogenesis-entering cell, i.e. the presence of crossing-over analogous to the one proceeding at meiosis. The found combinatorial process and the earlier-obtained data confirm the hypothesis on multi-dimensionality of inherited information coding. Differential polyteny of certain chromosome regions can lead to differences among homozygous plants having the same alleles in genes located in polytenized regions and controlling morpho-physiological traits.

## INTRODUCTION

Manifestation of the laws of inheritance discovered by G. Mendel is provided by two combinatorial processes, the first of which consists in a decrease of inherited factors number during sexual cells (gametes) formation, the second one – in inherited factors integration at a random contact and fusion of male and female gametes [1]. A decrease of inherited factors number, that proceeds due to a decrease of chromosome number during meiosis, is realized randomly and equi-probably, which may be designated as meiotic combinatorial process. The second combinatorial process, provided by a gametes random meeting and fusion, may be designated as pre-zygotic.

The meiotic combinatorial process takes place also in the agamospermous (apomictic, subsexual) way of plant seed reproduction realized without gametes genomes integration. The agamospermous progeny developing from egg cells is polymorphous if a mother plant was heterozygous on genes controlling marker traits [2, 3].

However, the detection of polymorphism in agamospermous progenies formed from somatic cells by means of mitotic agamospermy allowed us to conclude on the existence of one more combinatorial process [4]. It was hypothesized that this combinatorial process is provided by chromosome polytenization in mother plant cells and a random equi-probable diminution (loss) of an excess of chromatin by an embryogenesis-entering cell [5–7]. The frequency of this or that allele in such an agamospermous progeny is determined by a relative number of chromatides carrying this allele in mother plant cells.

It is known that polyteny is possible not only in somatic cells, but in various embryo sac cells [8–12]. Moreover, it was shown that the ratio of phenotypic classes in agamospermous progenies can be determined by a level of polyteny that occurs in a matured egg cell at the absence of pollination and before its entering embryogenesis [13]. Such egg cells entering embryogenesis is connected with the diminution (loss) of chromatin excess, which proceeds randomly, equi-probably, i.e. combinatorially. We designated this combinatorial process as post-meiotic apozygotic [13].

Finding the phenomenon allowed us to raise a question about the possibility regarding the existence of an analogous combinatorial process also in the sexual progeny formation. To solve this point, reliable and convenient marker traits, among which isozymes are the most available and simple, are necessary. Being different in their electrophoretic mobility, isozymes have a co-dominant manifestation character in hybrids, which allows one, on isozyme spectra, to differentiate clearly both crossed forms and a hybrid from each of parent forms [14, 15]. To reveal the combinatorial process in sexual reproduction, a cross of one red table beet (RT) plant and one green sugar beet (SB) plant was made [16]. These plants were homozygous on the genes controlling glucosephosphate isomerase (GPI2) and malic enzyme (ME1), but they were different in the electrophoretic mobility of these enzymes.

Heterozygous spectrum GPI2 was found in all the investigated 37 hybrid seeds (RT x SB) set in a red table beet plant. But the heterozygous spectrum of malic enzyme was found only in 36 seeds as one seed had its malic enzyme spectrum presented only by the maternal isozyme having fast electrophoretic mobility (phenotype FF). This fact was explained by the thing that the egg cell introduced allele *Me1-F* with a high polyteny level into the zygote. It was supposed that the polyteny of allele *Me1-F* led to combinatorial diminution on hypergeometrical probability distribution laws. As a result of this combinatorial process, the introduced male paternal allele *Me1-S* was substituted by mother allele *Me1-F* [16]. However, this conclusion required an additional proof as the phenotype, similar with the maternal one, is possible to appear not only as a result of the substitution of the paternal allele, but also in its inactivation. In case of inactivation of the paternal allele, the seed that reminds of a homozygote in its phenotype is the heterozygote carrying the null allele. Allelic inactivation of the enzyme locus may proceed in many ways; therefore, it cannot be a weighty proof of the combinatorial process. Only plant seed homozygosity on the gene controlling one marker enzyme, under its heterozygosity on the second marker, that is suggestive of its passed hybridization, can point out the existence of the combinatorial process. The present contribution is devoted to the finding out of this point and the choice between the two possible mechanisms.

MATERIALS AND METHODS

RT x SB hybrid seeds obtained from crossing red table beet (RT) N 19 and the green plant of inbred sugar beet line KWS-9c (SB) were involved in the study. The table beet plant had genotype *Gpi2-F/Gpi2-F* (short – *FF*) and *Me1-F/Me1-F* (short – *FF*) and, respectively, fast-migrating glucosephospate isomerase isozyme (GPI2, E.C. 5.3.1.9.) and malic enzyme (ME1, E.C. 1.1.1.40.). The sugar beet plant had genotype *Gpi2-S/Gpi2-S* (short – *SS*) and *Me1-S/Me1-S* (short – *SS*) and, respectively, slow-migrating GPI2 and ME1.

These two plants hybridization process was realized in 2008 by free cross-pollination in an isolated plot at the distance not less than two kilometers away from any other flowering beet plants. The obtained seeds of RT x SB hybrid had been sown in a hydroponic greenhouse in February, 2012, where the plants were then being grown during three months.

RT x SB plant phenotypes were determined by the analysis of GPI2 and ME1 isozymes in leaves and leafstalks. One of the hybrid plants flowered during growth. It allowed us to determine not only its phenotype, but the genotype by genetic analysis studying the seed progeny obtained from crossing it with test plants in summer, 2012.

Electrophoretic analysis of the vegetating plant phenotypes and the seed progeny obtained in analyzing crossings was carried out on the earlier-described methods [17].

RESULTS AND DISCUSSION

Forty plants were grown from the RT x SB hybrid seeds at our disposal, out of which 38 had their hybrid phenotype revealed on malic enzyme (ME1) and 2 plants (№ 18 and № 28) had the phenotype on ME1 similar with the homozygous mother phenotype. The appearance of these two plants is aberration of $F_1$ hybrid uniformity law. Plant phenotypes on GPI2 were determined obscurely at the 2-3-month growth stage. In 3 months of growth in the hydroponic greenhouse, the plants were planted in the open ground for further growth and vernalization. Out of two plants that had mother phenotype on ME1, one flowered without preliminary vernalization and that allowed us to analyze its genotype on the progeny. The phenotype on ME1 revealed in this plant could be conditioned either by genotype *FF* or *F0*. It is possible to differentiate these two states by, e.g., crossing this plant with the testing plant of genotype *SS*. Suppose that the studied plant has genotype *FF*, then, in cross *SS* Ч *FF*, a uniform progeny of heterozygous phenotype FS is to form. If the studied plant has genotype *F0*, then, in testing cross *SS* x *F0*, there forms the progeny that consists of two geno- and phenotypic classes: common heterozygotes *FS* and heterozygotes *S0*, whose phenotype is similar with that of *SS*.

Test plant № 12 with genotype *Me1-S/Me1-S*, *Gpi2-F/Gpi2*-S was used in the analysis as mother plant. In the obtained progeny, all the analyzed 45 seeds had their phenotype of heterozygotes *Me1-F/Me1-S*. It was indicative of the thing that the analyzed plant № 18 had homozygous genotype *Me1-F/Me1-F*. When studying the same progeny on enzyme GPI2, three phenotypic classes were found in it at ratio 8FF : 25FS : 9SS (Table 1) which indicated heterozygosity of plant № 18 on gene *Gpi2*. One more test plant (№ 11) of genotype *Me1-F/Me1-S*, *Gpi2-F/Gpi2-S* was involved in the experiment. Two phenotypic classes on enzyme ME1: 30FF : 20FS an three phenotypic classes on enzyme GPI2 17FF : 24FS : 13SS (Table 1) formed in the progeny after the pollination of this plant with the pollen of analyzed plant № 18.

Table 1.

**Genotypic analysis of plant № 18**

|  | Analyzed plant № 18 ||||||
|  | ME1 ||| GPI2 |||
|  | FF | FS | SS | FF | FS | SS |
| Tester plant № 12 of genotype *Me1-S/Me1-S, Gpi2-F/Gpi2-S* | 0 | 45 | 0 | 8 | 25 | 9 |
| Tester plant № 11 of genotype *Me1-F/Me1-S, Gpi2-F/Gpi2-S* | 30 | 20 | 0 | 17 | 24 | 13 |

Summing up the crossing results, it possible to conclude that the analyzed plant № 18 had genotype *Me1-F/Me1-F*, *Gpi2-F/Gpi2-S*. Heterozygosity on gene *Gpi2* is suggestive of the thing that plant N 18 was produced as a result of red table beet and sugar beet hybridization. However, homozygosity of the same plant on gene *Me1* indicates a specific mechanism of variability and also the thing that the appearance of homozygous phenotype FF among $F_1$ hybrid heterozygotes is not connected with the possible silencing of allele *Me1-S*. The obtained results confirm our earlier preliminary conclusions on the thing that aberration of $F_1$ hybrids uniformity is conditioned by the combinatorial process connected with an equi-probable diminution (loss) of chromatin excess from the embryogenesis-entering cell [16].

As this process was revealed in the zygote, it is possible to designate it as *zygotic combinatorial process*.

The analogous facts were found out earlier in the genetic analysis of agamospermous sugar beet progenies when plants, first considered as heterozygotes carrying, along with the normal enzyme locus allele, also the inactivated allele, were, actually, homozygotes on the normal allele [18, 19].

The data we obtained well agree to the known results of genetic investigations being carried out in animals. Thus, for example, there were experimentally bred mice that inherited both copies of mother chromosome 11 and none of the paternal copies [20].

When studying inherited human diseases, there was found the phenomenon known as single-parent disomy: when a child has two similar chromosomes 15 but both inherited from the father or from the mother [21]. Such situations may happen both in mice and human when there is excessive chromatin generation of some chromosome being accompanied by its division into two independent identical chromosomes and by the following random release of chromatin excess from the zygotic cell proceeding so that two identical chromosomes remain in the cell and the third chromosome, not similar with them, is lost.

This research is a logical completion of a series of contributions that began with bringing forth the hypothesis on multidimensionality of inherited information coding in plants [5–7] and were continued by obtaining genetic proofs of this hypothesis – first in agamospermous (subsexual) [22] and then in gamospermous (sexual) progenies [16]. This allows us to make some general conclusions regarding the peculiarities of zygotic and apozygotic combinatorial processes in plants, also to understand many biological phenomena explainable on the base of the proposed outlook on these processes.

1. Sugar beet was used as a model object to study variability in sexual (gamospermous) and subsexual (agamospermous) progenies. The agamospermous progenies of diploid and triploid sugar beet plants were diploid. The sequence of events underlying agamospermous reproduction can be briefly outlined the following way: separate chromosome regions or whole chromosome polytenization in egg cells, embryo sac cells or somatic nucellus and integuments cells; random equi-probable attachment of two out of a multitude of allelic gene copies (or chromatides) present in the nucleus to the nuclear membrane of the cell capable of agamospermous development; cell transition to embryogenesis and the onset of embryogenesis; chromosome duplication anticipating nucleus and cell division; first embryogenetic division and diminution (elimination) of allelic gene copies or whole chromatides unattached to the nuclear membrane. An important moment determining the genotype of a developing embryo is random equi-probable attachment of allelic gene copies to the nuclear membrane realized on one copy from each homological chromosome. Attachment of allelic copies to the nuclear membrane is described by hypergeometrical probability distribution formulae. It means that the ratio of phenotypic classes in agamospermous progenies is determined, for homozygotes, by the number of combinations – two from the present number of copies of the corresponding allele. For example, if one allele is presented by $n$ copies and the other by $m$ copies in a diploid cell, then the portion of homozygotes (homoallelic genotypes) on the first allele is determined according to the formula: $C^2_n/C^2_{n+m}$, on the second – $C^2_m/C^2_{n+m}$. Accordingly, the portion of heterozygous (heteroallelic) genotype is determined on the formula $C^1_n C^1_m / C^2_{n+m}$. The analogous combinatorial process is possible also in the sexual reproduction process if the alleles introduced to the zygote are presented by more than one copy.

2. The combinatorial process manifests itself only under mutual exchange of chromatides between chromosomes. Hence, in an embryogenesis-entering cell, there proceeds the crossing-over, analogous to the one which takes place at meiosis.

3. Manifestation of only mother allele of locus *Me1* in hybrid RT x SB is indicative of the thing that its polyteny degree in the zygote, before its entering embryogenesis, was higher than that of the introduced paternal allele. This well agrees to the thing that the frequency of polyteny in cells of female generative sphere is many times higher than that of the male [23].

4. Polytenization of different genes alleles proceeds independently and it is conditioned not only by these genes different chromosome localization, but by the presence of a multitude of independent reduplication onset points in each plant chromosome. It is this that explains our earlier-obtained data on the differences, revealed in agamospermous progenies, in the phenotypic classes ratios of enzymes controlled by non-allelic genes [4].

5. Revealing the combinatorial effect on locus *Me1* in only one of reciprocal hybrids, namely in hybrid RT x SB and the absence of such effect in hybrid SB x RT [16] points out the non-univocality of the processes that determine separate chromosome regions polytenization. One can see that it is not necessary that the allele belonging to the mother genome leads to its polytenization. It is not excluded that there are differences among alleles in their predisposition to polytenization.

6. The zygotic combinatorial process in the hybrid obtained from crossing beet forms of different origin (red table beet and sugar beet) can be considered as one of the mechanisms underlying hybrid disgenesis. An important trait of hybrid disgenesis is its non-reciprocality [24, 25] which we observe in the comparison of hybrids SB x RT and RT x SB. One of the commonly accepted explanations for hybrid disgenesis is the presence of mobile elements that change their location in a genome and lead to a change of closely located genes manifestation [26]. The other explanation, which is most similar with our outlook on genome structural-functional organization, was presented by J. Sved [27]. He explained hybrid disgenesis in terms of chromosome spatial organization in the cell nucleus. He believed that hybrid disgenesis in drosophila is manifested in case when a chromosome (or chromosomes) of paternal line lacks information to be strictly organized and be located in the zygote nucleus whose nuclear membrane forms from the egg cell nucleus [27]. According to our standpoint, the impossibility of organization in the zygote nucleus for paternal chromosomes may be explained by a spatial limitation conditioned by differences in the polyteny degree of chromosomes introduced in a zygote by the egg cell and male gamete. We earlier mentioned the thing that spatial interrelations between parent chromosomes in a zygote may be the reason for the changes of mobile elements locations, which have to give space to larger chromosome regions [6, 7]. Spatial relations are determined by differences in the degree of chromosome polyteny in crossed forms and manifest themselves in chromosome competition to contact the nucleus membrane which, finally, leads to a change of mobile elements locations and to, connected with it, numerous aberrations in hybrid disgenesis. One can see that, according to this outlook, mobile elements moving is just a consequence, but not the reason for variability, i.e. polyteny, but not mobile elements, can be considered the main variability 'engine' in hybrid disgenesis.

One may suppose that hybrid disgenesis appears not only in some gene expression disturbance but also in the loss of offspring viability and decrease of seed set.

7. The combinatorial process expressed at a certain frequency is indicative of the thing that one and the same allele, as a consequence of polyteny, can be presented by a different copies number in different plants. Hence, plants of homozygous line, having similar nucleotide sequences, nevertheless, can differ from each other in the dose of one and the same homozygous genes. Differences in genes dosage among the plants of one line are the base of their phenotypic distinctions. These conclusions well agree to the viewpoints on multi-dimensionality of inherited information coding [5–7], according to which polyteny is considered as coding in the second dimension. In this light, the affinity of nucleotide sequences (coding in the first dimension) in plants of a high-inbred line is not sufficient for these plants classification as highly identical, as, herein, differences in the degree of polyteny are neglected. This standpoint can be the ground for the possibility of carrying out selections in pure lines and it contradicts to Johannsen's viewpoint rejecting the possibility of selection in pure lines [28].

8. The viewpoints on polyteny, diminution and combinatorial process allow us to hypothesize the mechanism underlying the phenomenon of '*penetrance*' [29]. 37 $F_1$ RT x SB hybrid plants were investigated in our experiments at the seed stage [16] and 40 plants, in this experiment, – at the vegetation stage. Out of all 77 $F_1$ plants, 74 plants had their heterozygous phenotype on malic enzyme (ME1) and 3 had homozygous phenotype on this enzyme. Hence, the plant share of heterozygous phenotype, in which paternal allele *Me1-S* was expressed, is equal to 74/77 = 0,96. This value can be considered as the penetrance value of allele *Me1-S*.

Analogously, it is possible to explain also such a widely known phenomenon as nucleolar dominance [30]. The impossibility of a precise prediction and answer to the question about the thing which parent nucleolus will be expressed in the hybrid is explained, as we believe, by the combinatorial process that, as is known, is probable. One can suppose the presence of a high polyteny level of both parents nucleolus-coding chromosome regions leading to a random equi-probable diminution (loss) of chromatin excess in an embryogenesis-entering zygote.

9. The frequency of combinatorial processes in sexual progenies is considerably lower than that in the agamospermous. It points out the existence the specific mechanisms that hinder polyteny in gametes and the absence of such mechanisms in cells capable of passing on to embryogenesis by means of agamospermy. Agamospermous reproduction may be accompanied by incomplete diminution of chromatin excess and an increase of certain genes dosage [22]. It is this thing that can explain some increase of DNA content in cell nuclei of agamospermous progenies estimated both cytometrically and on the number of chloroplasts in stomata guard cells [31, 32].

10. The initial red table and sugar beet plants involved in hybridization were biannual and, hence, they had homozygous phenotype on recessive allele *b* that determines this phenotype. Spontaneous appearance of annuals among biannuals is far from being frequent [33]. The zygotic combinatorial process, as it can be seen from the obtained data, is also not a frequent phenomenon. Therefore, simultaneous manifestation of two relatively rare events – zygotic combinatorial process and a change of developmental type from bi- to annual – is, obviously, not random. This fact can be explained having supposed that gene *b/b* controlling the biannual developmental type was polytenized the same way as enzyme locus *Me1*, and combinatorial process was also proceeding among allelic copies of gene *b/b*. But then there is the other question: if the initial plants had a biannual developmental type, then how could the dominant phenotype controlled by allele *B* appear in the progeny? It is well known from literary sources that the presence of multiple gene copies leads to its instability [34, 35]. Therefore, appearance of the dominant allele can be in favor of the supposition on the presence of polyteny in the locus controlling one and two-year developmental types.

We also revealed instability in enzyme loci in cases when the phenotypic classes ratio in investigated agamospermous triploid plant progenies was indicative of polyteny state of enzyme genes [22]. Appearance of the phenotypes corresponding to the alleles that were absent in the initial mother plant [36] in agamospermous progenies produced from diploid heterozygous plants can also be an example. In this experiment, simultaneous occurrence of two processes (change of plant developmental type and combinatorial process in chromosome regions carrying locus *Me1*) may be conditioned not only by polyteny of genes *Me1* and *B/b*, but also either their close linkage or close arrangement in the spatial structure of the nucleus.

Thus, we obtained additional proofs of the thing that aberration of the Mendel ratios in sexual progenies can be explained by polyteny of chromosome regions carrying marker genes and by excessive chromatin copies diminution from the embryogenesis-entering zygote. This well agrees to the thing that its on the base of these viewpoints that it becomes possible to explain phenotypic ratios in most of the sugar beet progenies, produced in agamospermous way, we analyzed earlier.

The evolutionary meaning of differential polyteny and combinatorial processes, and diminution connected with them consists not only in their influence on phenotypic ratios in further generations, but also in the thing that polyteny, just as the whole process of DNA synthesis, depends on a big number of internal and external factors [37–42]. Therefore, differential polyteny can be considered as a way of recording information about acquired traits.

## REFERENCES

1. *Mendel G.* Versuche uber Pflanzen-Hybriden. Verhandlungen des naturforschenden Vereines in Brunn (Abhandlungen). 1866. Bd. 4. S. 3–47.

2. *Gustafsson A.* Apomixis in higher plants. I-III. Lunds. Univ. Arsskr. N.F. Adv. 1946-1947. 42-43: 1–370.

3. *Maletskii S.I.,Sukhareva N.B., Baturin S.O.* Sex inheritance in apomictic seedlings of garden strawberry (*Fragaria* x *ananassa* Duch.). Genetika. 1994. V.30. N 2. P. 237–243 (In Russian).

4. *Levites E.V., Shkutnik T., Ovechkina O.N., Maletskii S.I.* Pseudosegregation in the agamospermic progeny of male sterile plants of the sugar beet (*Beta vulgaris* L.). Doklady Akademii Nauk. V. 362. N 3. P. 430–432 (In Russian).

5. *Levites E.V.* Sugarbeet plants produced by agamospermy as a model for studying genome structure and function in higher plants. Sugar Tech. 2005. V. 7. № 2–3. P. 67–70.

6. *Levites E.V.* Marker enzyme phenotype ratios in agamospermous sugarbeet progenies as a demonstration of multidimensional encoding of inherited information in plants. 2007. http://arxiv.org/abs/q-bio/0701027

7. *Levites E.V.* Sugar beet as a model object in the investigation of plant inherited information coding. Encyclopaedia of genus Beta. Beet biology, genetics and breeding. (Col. of sci. papers). Novosibirsk: "Sova" publ., 2010. P. 302–317. (In Russian)

8. *Tschermak-Woess E.* Uber Kernstrukturen in den endopolyploiden Antipoden von *Clivia miniata.* Chromosoma. 1957. Bd. 8. S. 637–649.

9. *Hasitschka-Jenschke G.* Vergleichende karyologische untersuchungen an antipoden. Chromosoma. 1959. Bd. 10. S. 229–267.

10. *Ivanovskaya E.V.* Functional embryology of wheat antipodes polytene chromosomes. Cytology. 1973. V. 15. № 12. P. 1445–1452. (In Russian).

11. *Nagl W.* Polytene chromosomes of plants. Int. Rev. Cytol. 1981. V. 73. P. 21–53.

12. *Carvalheira G.* Plant polytene chromosomes. Genet. Mol. Biol. 2000. V. 23. № 4. P. 1043–1050.

13. *Levites E.V., Kirikovich S.S.* Post-meiotic apozygotic combinatory process in sugar beet (*Beta vulgaris* L.). Advances in Bioscience and Biotechnology. 2012. V. 3. № 1. P. 75–79.

14. *Schwartz D.* Genetic studies on mutant enzymes in maize. Synthesis of hybrid enzymes by heterozygotes. Proc. Natl. Acad. Sci. USA. 1960. V. 46. № 9. P. 1210–1215.

15. *Scandalios J.G.* Genetic control of multiple forms of enzymes in plants: a review. Biochem. Genet. 1969. V. 3. № 1. P. 37–79.

16. *Levites E.V.* Violation of the low of uniformity of the first generation of hybrids. Russ. J. Genetics. 2012. V. 48. № 11. P. 1158–1161.

17. *Meizel S.*, *Markert C.L.* Malate dehydrogenase isozymes of the marine snail *Ilyanassa obsoleta*. Arch. Biochem. Biophys.1967. V. 122. P. 753–765.

18. *Levites E.V.* Redetermination: an interesting epigenetic phenomenon associated with mitotic agamospermy in sugar beet. Sugar Tech. 2002. V. 4. N 3&4. P. 137-141.

19. *Levites E.V., Kirikovich S.S.* Epigenetic variability of unlinked enzyme genes in agamospermous progeny of sugar beet. Sugar Tech. 2003. V. 5. № 1/2. P. 57-59.

20. *Cattanach B.M., Kirk M.* Differential activity of maternally and paternally derived chromosome regions in mice. Nature. 1985. V. 315. P. 496–498.

21. *Knoll J.H., Nicholls R.D., Magenis R.E. et al.* Angelman and Prader-Willi syndromes share a common chromosome 15 deletion but differ in parental origin of the deletion. Am. J. Med. Genet. 1989. V.32. № 2. P. 285–290.

22. *Levites E.V., Kirikovich S.S.* Autosegregation of enzyme loci in agamospermous progenies of triploid plants of sugar beet (*Beta vulgaris* L.). Russ. J. Genetics. 2011. V. 47. № 7. P. 836–841.

23. *Kojima A., Nagato Y.* Diplosporous embryo-sac formation and the degree of diplospory in *Allium tuberosum*. Biomedical and Life Sciences. 1992. V. 5. № 1. P. 72–78.

24. *Kidwell M.G., Kidwell J.F.* Cytoplasm-chromosome interactions in *Drosophila melanogaster*. Nature. 1975. V. 253. P. 755–756.

25. *Kidwell M.G., Kidwell J.F., Sved J.A.* Hybrid dysgenesis in *Drosophila melanogaster*: a syndrome of aberrant traits including mutation, sterility and male recombination // Genetics. 1977. V. 86. P. 813–833.

26. *Engels W.R. P*-element in *Drosophila melanogaster*. Mobile DNA /edited by D.E. Berg and M.M. Howe/. American Society for Microbiology. Washington, DC, 1989. P. 437–484.

27. *Sved J.A.* Hybrid dysgenesis in *Drosophila melanogaster*: a possible explanation in terms of spatial organization of chromosomes. Aust. J. Biol. Sci. 1976. V. 29. P. 375–388.

28. *Johannsen W.* Ueber Erblichkeit in Populationen und reinen Linien. Eine Beitrag zur Beleuchtung schwebender Selektionsfragen. Jena: Gustav Fischer, 1903. P. 58–59.

29. *Timofeev-Risovsky N.V.* About phenotypic manifestation of genotype. 1. Gene variations in radius incompletes of *Drosophila funebris*. Exper. Biolog. j. Series A. 1925. V. 1. I. 3/4. P. 93–142.

30. *Navashin M.* Chromosome alteration caused by hybridisation and their bearing upon certain general genetic problems. Cytologia. 1934. V.5. № 2. P.169–203.

31. *Matzk F., Hammer K., Schubert I.* Co-evolution of apomixis and genome size within the genus *Hypericum*. Sex. Plant Report. 2003. V. 16. P. 51-58.

32. *Yudanova S.S., Maketskaya E.I., Maletskii S.I.* Epiplastome variation of the number of chloroplasts in stomata guard cells of sugar beet (*Beta vulgaris* L.). Russ. J. Genetics. 2004. V. 40. № 7.

33. Unpublished date

34. *Gordenin D., Resnick M.A.* Yeast ARMs (DNA at-risk motifs) can reveal sources of genome instability // Mutat. Res. 1998. V.400. P. 45–58.

35. *Kim H.-M.* Genome instability induced by triplex forming mirror repeats in *S.cerevisiae*. In partial fulfillment of the requirements for the degree Doctor of Philosophy in the School of Biology. Georgia Institute of technology, 2009. 134 p.

36. *Levites E.V., Shkutnik T., Shavorskaya O.A., Denisova F.Sh.* Epigenetic variability in agamospermous progeny of sugar beet. Sugar Tech. 2001. V. 3. № 3. P. 101–105.

37. *Evans G.M.* Nuclear changes in flax. Heredity. 1968. V. 23. P. 25–38.

38. *Nagl W.* Photoperiodic control of activity of the suspensor polytene chromosomes in *Phaseolus vulgaris*. Z. Pflanzenphysiol. 1973. V. 70. P. 350–357.

39. *Durrant A., Timmis J.N.* Genetic control of environmentally induced changes in *Linum*. Heredity. 1973. V. 30. No 3. P. 369–379.

40. *Cullis C.A.* DNA differences between flax genotrophs. Nature. 1973. V. 243. P. 515–516.

41. *Zhimulev I.F.* Polytene chromosomes: morphology and structure. Novosibirsk: Nauka. Sib. Branch. 1992. 480 P. (In Russian).

42. *Kirikovich S.S.* Epigenetic variability of enzyme loci in sugar beet (*Beta vulgaris* L.) under many-factor influences. PhD thesis. Tomsk, 2004. 129 p. (In Russian).